\documentclass[twocolumn,amsmath,amssymb]
   {revtex4}
\usepackage{graphicx}
\usepackage{dcolumn}
\usepackage{bm}

\begin{document}
\title[Wormholes supported by Chaplygin gas]
{Theoretical construction of wormholes supported by Chaplygin gas}
\author{Peter K. F. Kuhfittig}
\address{Department of Mathematics\\
Milwaukee School of Engineering\\
Milwaukee, Wisconsin 53202-3109 USA}
\date{\today}

\begin{abstract}\noindent
The purpose of this paper is to examine the possible existence or
construction of traversable wormholes supported by generalized 
Chaplygin gas (GCG) by starting with a general line element and the 
Einstein tensor, together with the equation of state, thereby 
continuing an earlier study by the author of wormholes supported by 
phantom energy.  Numerical techniques are used to demonstrate the 
existence of wormhole spacetimes that (1) meet the flare-out 
conditions at the throat, (2) are traversable by humanoid travelers,
thanks to low tidal forces and short proper distances near the throat, 
and (3) are asymptotically flat.  There appears to be an abundance of 
solutions that avoid an event horizon, suggesting the possibility 
of naturally occurring wormholes.  
\end{abstract}

\maketitle 
MOS (AMS) Subject Classification.  83C05

\phantom{a}

\section{Introduction}\noindent
Interest in traversable wormholes, initiated by Morris and Thorne~ 
\cite{MT88}, has been greatly renewed in part because of the discovery 
that our Universe is undergoing an accelerated expansion \cite{aR98, 
sP99}.  In other words, $\overset{..}a>0$ in the Friedmann equation 
$\overset{..}{a}/a=-\frac{4\pi}{3}(\rho+3p)$, using units in which 
$G=c=1$.  The cause of this acceleration is taken to be a negative 
pressure \emph{dark energy} with equation of state $p=-K\rho$, $K>\frac
{1}{3}$, and $\rho>0$, where $K$ is a constant, $p$ the spatially 
homogeneous pressure, and $\rho$ the energy density.  Of particular 
interest is the case $K>1$, referred to as \emph{phantom energy},
since it leads to a violation of the null energy condition, an 
essential requirement for maintaining a wormhole \cite{MT88}.  Matter 
violating the null energy condition is usually called \emph{exotic}.  
Since the notion of dark or phantom energy ordinarily applies only 
to a homogeneous distribution of matter, phantom energy is not 
automatically a candidate for exotic matter.  Fortunately, the 
extension to spherically symmetric homogeneous spacetimes has been 
carried out \cite{SK04}.  

An alternative model is based on Chaplygin gas, whose equation 
of state is given by 
$p=-\frac{A}{\rho}$.  Another possibility is generalized Chaplygin 
gas (GCG), whose equation of state is $p=-\frac{A}{\rho^{\alpha}}$, 
$0<\alpha\le 1$\, \cite{KMP01, BBS02}. Cosmologists became 
interested in this form of matter when it turned out to be a 
candidate for unifying dark matter and dark energy.  To see this, 
consider the energy conservation equation 
$\overset{.}{\rho}=-3\overset{.}{a}(\rho+p)/a$ in a flat FRW 
spacetime and substitute the equation of state 
$p=-\frac{A}{\rho^{\alpha}}$.  The result is 
\begin{equation}\label{E:FRW}
   \rho=\left(A+\frac{B}{a^{3(1+\alpha)}}\right)^{1/(1+\alpha)},
\end{equation}
where $B$ is a constant of integration.  It is now seen that 
$\rho\sim a^{-3}$ at early times, that is, $\rho$ behaves like matter,
while in later times like a cosmological constant ($\rho=\text
{constant}$).  Previous models required two distinct fields, one to 
describe dark matter and the other dark energy, but one can argue, 
as in Ref.~\cite{BTV02a}, that these ought to be different 
manifestations of the same entity.  One possible motivation for this 
model came from the field theory points of view \cite{BTV02b}.  
Another interesting possibility is discussed in Ref.~\cite{FGdS02}:
starting with the Nambu-Goto action of string theory, the Chaplygin 
gas appears after considering $d$-branes in a $(d+2)$-dimensional 
spacetime.  Another attraction is that it admits a supersymmetric 
extension.

Observationally, the GCG model has not been without its problems.  
As noted in Ref.~\cite{BBS04}, while the model has successfully 
withstood various phenomenological tests over several years, there 
is some concern that it produces unphysical oscillations in the 
matter power spectrum.  It is shown in Ref.~\cite{BBS04}, however, 
that these problems can be circumvented.  Furthermore, very recent 
studies \cite{ZF05, BTV06} have concluded that the earlier criticisms 
were based on the oversimplifying assumption of an adiabatic cosmic 
medium.

On the other hand, according to another recent study (Ref.~ 
\cite{ZWZ06}), the special case $\alpha=1$, corresponding to the 
original Chaplygin gas, may very well have to be excluded.  One should 
therefore concentrate on the GCG case by keeping $\alpha<1$.

In a recent paper \cite{pK06b} the author made a systematic study of 
exact solutions of wormhole spacetimes supported by phantom energy 
by starting with the general line element and equation of state.  It 
is shown that there are only two ways to insert the redshift function 
``by hand."  Doing so leads to the exact solutions in Refs.~ 
\cite{fL05} and \cite{oZ05}.  Assigning a specific function to $\rho$ 
leads to the exact solution in Ref.~\cite{sS05}.  Additional exact 
solutions that simultaneously avoid an event horizon are extremely 
rare.  Nevertheless two new solutions were found.  Included in the 
discussion are the junction conditions for matching each solution 
to an exterior Schwarzschild solution, as well as traversability 
criteria.

Ref.~\cite{fL06} makes use of specific redshift and shape functions 
in the manner of the original Morris and Thorne paper.  In this paper 
we continue in the spirit of Ref.~\cite{pK06b}, that is, we start 
with the general line element and the Einstein tensor, together with 
the equation of state, and continue the analysis while avoiding, as 
much as possible, assigning specific functions to the metric 
coefficients.

\section{The problem}\noindent
Consider the general line element for describing a wormhole:
\begin{equation}\label{E:line1}
  ds^2=-e^{2\Phi(r)}dt^2+e^{2\Lambda(r)}dr^2
      +r^2(d\theta^2+\text{sin}^2\theta\,d\phi^2).
\end{equation}
In this context $\Phi(r)$ is called the \emph{redshift function};
for this function, $e^{2\Phi(r)}$ must never vanish to avoid an 
event horizon.  $\Lambda(r)$ is related to the \emph{shape function} 
$b(r)=r(1-e^{-2\Lambda(r)})$, i.e., $e^{2\Lambda(r)}=1/[1-b(r)/r]$.  
The shape function determines the spatial shape of the wormhole 
as viewed, for example, in an embedding diagram.  By the very 
definition of wormhole, if the throat is at $r=r_0$, then 
$b(r_0)=r_0$.  As a consequence, $\Lambda(r)$ has a vertical 
asymptote at $r=r_0$: $\lim_{r \to r_0+}\Lambda(r)=+\infty$.  
To obtain a traversable wormhole, the shape function must obey the 
usual flare-out conditions at the throat \cite{MT88}: $b'(r_0)<1$ 
and $b(r)<r$; also required is asymptotic flatness, i.e., 
$b(r)/r\rightarrow 0$ as $r\rightarrow\infty$.    

The next step is to list the components of the Einstein tensor 
in the orthonormal frame \cite{pK06b}:
\begin{equation}\label{E:Einstein1}
  G_{\hat{t}\hat{t}}=\frac{2}{r}e^{-2\Lambda(r)}\Lambda'(r)
    +\frac{1}{r^2}\left(1-e^{-2\Lambda(r)}\right),
\end{equation}
\begin{equation}\label{E:Einstein2}
   G_{\hat{r}\hat{r}}=\frac{2}{r}e^{-2\Lambda(r)}\Phi'(r)
   -\frac{1}{r^2}\left(1-e^{-2\Lambda(r)}\right),
\end{equation}
\begin{multline}\label{E:Einstein3}
   G_{\hat{\theta}\hat{\theta}}=G_{\hat{\phi}\hat{\phi}}\\
   =e^{-2\Lambda(r)}\left(\Phi''(r)-\Phi'(r)\Lambda'(r)+
   [\Phi'(r)]^2\phantom{\frac{1}{r}}\right.\\
    \left. +\frac{1}{r}\Phi'(r)-\frac{1}{r}\Lambda'(r)\right).
\end{multline}
Since the Einstein field equations $G_{\hat{\alpha}\hat{\beta}}
=8\pi T_{\hat{\alpha}\hat{\beta}}$ imply that the stress-energy 
tensor $T_{\hat{\alpha}\hat{\beta}}$ is proportional to the Einstein 
tensor, the only nonzero components are $T_{\hat{t}\hat{t}}=\rho(r)$, 
$T_{\hat{r}\hat{r}}=p(r)$, and $T_{\hat{\theta}\hat{\theta}}
=T_{\hat{\phi}\hat{\phi}}=p_t(r)$, the transverse pressure.  
From the Einstein field equations and the equation of state 
$p=-\frac{A}{\rho^{\alpha}}$, we obtain
 $G_{\hat{t}\hat{t}}=8\pi\rho$ and 
$G_{\hat{r}\hat{r}}=8\pi(-A\rho^{-\alpha})$, yielding the 
following equation: 
\begin{multline}\label{E:basiceq1}
  \frac{1}{8\pi}\left[\frac{2}{r}e^{-2\Lambda(r)}\Phi'(r)
   -\frac{1}{r^2}\left(1-e^{-2\Lambda(r)}\right)\right]\\
   =-\frac{A}{\left\{\frac{1}{8\pi}\left[\frac{2}{r}e^{-2\Lambda(r)}
    \Lambda'(r)+\frac{1}{r^2}\left(1-e^{-2\Lambda(r)}\right)
     \right]\right\}^{\alpha}}.
\end{multline}
Unlike the phantom energy case $(p=-K\rho)$, the equation of state 
has the form of a quotient.  The resulting differential equation 
(\ref{E:basiceq1}) does not have an obvious exact solution, so 
that the analysis depends on numerical/graphical techniques.

Recall next the null energy condition, which requires the 
stress-energy tensor $T_{\alpha\beta}$ to obey $T_{\alpha\beta}
\mu^{\alpha}\mu^{\beta}\ge0$ for all null vectors.  In our
orthonormal frame, for $(\mu^{\hat{t}},\mu^{\hat{r}},0,0)=(1,1,0,0)$,
a radial outgoing null vector, the condition becomes $T_{\hat{t}\hat{t}}
+T_{\hat{r}\hat{r}}=\rho+p\ge0$. Wormholes must necessarily violate
this condition at the throat \cite{MT88}. Since 
$\lim_{r \to r_0+}\Lambda(r)=+\infty$, Eq.~(\ref{E:basiceq1}) now 
implies that 
\[
   \frac{1}{8\pi}\left(-\frac{1}{r_0^2}\right)+\frac{A}
     {\left(\frac{1}{8\pi}\right)^{\alpha}\left(\frac{1}
      {r_0^2}\right)^{\alpha}}=0.
\]
Solving for $A$, we have $A=1/(8\pi r_0^2)^{\alpha+1}$.  We will 
see below that we actually need
\begin{equation}\label{E:constraint1}
   A<\frac{1}{(8\pi r_0^2)^{\alpha+1}}.
\end{equation}
Using the above equation of state,
\begin{equation}\label{E:constraint2}
   p=-\frac{A}{\rho^{\alpha}},\quad 0<\alpha\le1,
\end{equation}
in conjunction with $\rho+p<0$, yields the constraint 
\begin{equation}\label{E:constraint3}
   \rho<A^{1/(\alpha+1)},
\end{equation}
which, in turn, implies that $B<0$ in Eq~(\ref{E:FRW}).  (The 
constraints (\ref{E:constraint1}) and (\ref{E:constraint3}) are 
also discussed in Ref. \cite{fL06}.)

\section{Theoretical analysis}\noindent
As already noted, Eq.~(\ref{E:basiceq1}) does not lend itself to 
finding a simple exact solution, so that some numerical techniques 
will be needed.  That is the topic of  Sec.~ \ref{S:numerical}.  In 
this section we determine some general characteristics of GCG 
wormholes.   The analysis is based on the assumption that a solution 
$\Lambda(r)$ of Eq.~(\ref{E:basiceq1}) exists for a wide range of 
choices for $\Phi'(r)$ and such that $\lim_{r \to r_0+}
\Lambda(r)=+\infty$.  We also assume that $\Phi'(r)$ is continuous.

Suppose we rewrite Eq.~(\ref{E:basiceq1}) as follows:
\begin{multline}\label{E:basiceq2}
  \frac{(8\pi)^{1+1/\alpha}r(r^2)^{1/\alpha}A^{1/\alpha}}
   {\left(1-e^{-2\Lambda(r)}\right)\left[1-e^{-2\Lambda(r)}
     -2re^{-2\Lambda(r)}\Phi'(r)\right]^{1/\alpha}}\\
    =\frac{1}{r}+\frac{2e^{-2\Lambda(r)}\Lambda'(r)}
          {1-e^{-2\Lambda(r)}}.
\end{multline}
Now define the dimensionless function 
\begin{multline}\label{E:Form1}
  F(r)=\\ \int_{r_0}^r \frac{(8\pi)^{1+1/\alpha}r'[(r')^2]^{1/\alpha}A^{1/\alpha}dr'}
   {\left(1-e^{-2\Lambda(r')}\right)\left[1-e^{-2\Lambda(r')}
     -2r'e^{-2\Lambda(r')}\Phi'(r')\right]^{1/\alpha}}. 
\end{multline}
Observe that since $e^{-2\Lambda(r)}\rightarrow 0$ as $r\rightarrow 
r_0+$, the integrand is sectionally continuous.  This implies that 
$F(r)$ is defined and continuous on any interval $(r_0,r]$.  
Moreover, $F(r_0)=0$.

Returning to Eq.~(\ref{E:basiceq2}), integration yields
\[
   F(r)=\text{ln}\left(1-e^{-2\Lambda(r)}\right)+\text{ln}\,r
     +\text{ln}\,c.
\]
So
\begin{equation}\label{E:Form2}
  e^{F(r)}=c\,r\left(1-e^{-2\Lambda(r)}\right).
\end{equation}
Once again, $\lim_{r \to r_0+}\Lambda(r)=+\infty$ and $F(r_0)
=0$.  So by Eq.~(\ref{E:Form2}), $c=1/r_0$ and 
\begin{equation}\label{E:shape1}
   e^{2\Lambda(r)}=\frac{1}{1-\frac{r_0}{r}\,e^{F(r)}}.
\end{equation}
The line element then takes on the form
\[
  ds^2=-e^{2\Phi(r)}dt^2+\frac{dr^2}{1-\frac{r_0}{r}\,e^{F(r)}}+
        r^2(d\theta^2+\text{sin}^2\theta\,d\phi^2).
\]

The next step is to show that the flare-out conditions are satisfied 
at the throat.  Given that $b(r)=r\left(1-e^{-2\Lambda(r)}\right)$,
it follows that $b(r)<r$ for $r>r_0$.  Also, Eq.~(\ref{E:shape1}) 
implies that
\[
   b(r)=r\left(1-1+\frac{r_0}{r}e^{F(r)}\right)
\]
and
\[
     b(r)=r_0\,e^{F(r)}.
\]
By Eq.~(\ref{E:Form1}),
\begin{multline*}
b'(r)=r_0e^{F(r)}\times\\
   \frac{(8\pi)^{1+1/\alpha}r(r^2)^{1/\alpha}A^{1/\alpha}}
   {\left(1-e^{-2\Lambda(r)}\right)\left[1-e^{-2\Lambda(r)}
     -2re^{-2\Lambda(r)}\Phi'(r)\right]^{1/\alpha}}. 
\end{multline*}
If we assume that
\[
   A<\frac{1}{(8\pi r_0^2)^{\alpha+1}},
\]
then
\[
     A^{1/\alpha}<\frac{1}{(8\pi)^{1+1/\alpha}(r_0^2)^{1+1/\alpha}},
\]
whence $b'(r_0)<1$.

In summary, the analysis in this section depends primarily on the 
asymptotic behavior of $\Lambda(r)$ at the throat.  It has not been 
established that the existence of $\Lambda(r)$ yields a traversable 
wormhole.  In addition, its dependence on $\Phi(r)$ has to be closely 
examined, just as it is for phantom energy supported wormholes 
\cite{pK06b}.  These ideas are discussed in the next section.


\section{Numerical analysis}\label{S:numerical}\noindent
In this section we turn to the solution of Eq.~(\ref{E:basiceq2}), 
rewritten as follows:
\begin{multline}\label{E:basiceq3}
  \Lambda'(r)=\\ \frac{\frac{1}{2}(8\pi)^{1+1/\alpha}r(r^2)^{1/\alpha}
   A^{1/\alpha}e^{2\Lambda(r)}}{\left[1-e^{-2\Lambda(r)}
     -2re^{-2\Lambda(r)}\Phi'(r)\right]^{1/\alpha}}
       -\frac{1}{2r}\left(e^{2\Lambda(r)}-1\right). 
\end{multline} 
Here $A$ is taken to be less than $1/(8\pi r_0^2)
^{\alpha+1}$ and $\alpha$ to be close to unity.  (In fact, for 
convenience we let $\alpha=1$, at least for now.)  In order to 
generate numerical/graphical output, it is necessary to 
choose some specific value for the size of the throat (such as 
$r_0=2$) and to pick an initial value [such as $(2.000001,5)$] to 
simulate the asymptotic behavior.  As one might expect, 
qualitatively, the form of $\Lambda(r)$ remains roughly the same 
for a large range of $\Phi'$s.

\subsection{Regression and traversability}\noindent
In this section we turn to the question of traversability by 
humanoid travelers.  One concern is the calculation of the proper 
distance from the throat of the wormhole to a point outside:
\begin{equation}\label{E:proper1}
  \ell(r)=\int_{r_0}^r e^{\Lambda(r')}dr'.
\end{equation} 
For this calculation a specific $\Lambda(r)$ would be desirable.  
Regression equations from the numerical output are easy to obtain 
but tend to produce a poor fit---with one notable exception: 
$\Lambda(r)=a+b\,\text{ln}(r-r_0)$.  This form yields an excellent 
fit for any $\Phi(r)$ with small $|\Phi'(r)|$ and an almost perfect 
fit for $\Phi'(r)\equiv 0$.  ($\Phi'(r)$ can be either positive or 
negative; as $|\Phi'(r)|$ increases, 
$\Lambda(r)=a+b\,\text{ln}(r-r_0)$ yields an ever poorer fit.)
Some of the regression equations using $r_0=2$ with the 
corresponding $\Phi'$ are given next:
\begin{align*}
\Phi'(r)\equiv 0:\qquad &\Lambda(r)=1.20-0.23\,\text{ln}(r-2)\\
\Phi'(r)=\frac{10}{r^4}:\qquad  &\Lambda(r)=1.25-0.22\,\text{ln}(r-2)\\
\Phi'(r)=10e^{-r}:\qquad &\Lambda(r)=1.32-0.20\,\text{ln}(r-2)\\
\Phi'(r)=\frac{10}{r^2}:\qquad  &\Lambda(r)=1.45
      -0.18\,\text{ln}(r-2)\\
\Phi'(r)=\frac{10}{r^{1.5}}:\qquad   &\Lambda(r)=1.57-0.16\,\text{ln}(r-2)
\end{align*}
Observe that each $\Lambda(r)$ has a vertical asymptote at $r=2$ and that 
$\Lambda'(r)<0$ for $r>2$. The numerical solutions were obtained only in 
the vicinity of the throat.  The reason for this is discussed in 
Subsec. C.

\emph{Remark 1.} We may assume that the equations are fairly typical for 
GCG wormholes since there is nothing special about the choice of $r_0$ 
or the initial conditions.  Also implied is that the wormholes are 
macroscopic.

From the standpoint of traversability, a relatively small 
$|\Phi'(r)|$ results in relatively small tidal forces \cite{MT88}. 
In fact, for $\Phi'(r)\equiv 0$, the radial tidal force is zero.

The proper distance from the throat $r=r_0$ to a point outside, say 
$r=kr_0$ for some $k>0$, is also going to be relatively small thanks 
to the resulting ln-form.  As an example, for 
$\Lambda(r)=1.20-0.23\,\text{ln}(r-r_0)$, we have from 
Eq.~(\ref{E:proper1})
\begin{multline}\label{E:proper2}
  \int_{r_0}^{kr_0}e^{1.20-0.23\,\text{ln}(r-r_0)}dr\\
   =e^{1.20}\int_{r_0}^{kr_0}(r-r_0)^{-0.23}dr=
     e^{1.20}r_0^{0.77}\frac{(k-1)^{0.77}}{0.77},
\end{multline}  
corresponding to the coordinate distance $[r_0,kr_0]$.   

We conclude that a wormhole with small $|\Phi'(r)|$ in the vicinity of 
the throat is likely to be traversable in the sense of having low tidal 
forces and short proper distances resulting from the ln-form.  A 
wormhole is therefore traversable in this sense if, and only if, 
the numerical output for $\Lambda(r)$ closely fits the form 
$\Lambda(r)=a+b\,\text{ln}(r-r_0)$.

\emph{Remark 2.} Although of primary importance, $|\Phi'(r)|$ is not the 
only concern involving the redshift function.  Returning to line 
element (\ref{E:line1}), even if $\lim_{r \to +\infty}\Phi(r)=0$, a 
large $|\Phi(r)|$ near the throat is also undesirable: depending on the 
sign of $\Phi(r)$, clocks fixed at $r=r_0$ will either run much faster 
or much slower than clocks outside the wormhole.  So our definition 
of traversability does not include a study of proper traversal times.

\subsection{Other solutions}\noindent
One of the main conclusions in Ref.~\cite{pK06b} is the extreme difficulty in 
obtaining exactly solvable wormhole models without getting an unwanted event 
horizon.  While the question of exact solutions does not arise in this study, 
based on Eq.~(\ref{E:basiceq3}) and numerical trials, choosing an ``arbitrary" 
$\Phi'(r)$ would not ordinarily yield an event horizon.  The choices for 
$\Phi$ are therefore enormous.

Another fortunate circumstance is the occurrance of $\Lambda(r)=a+
b\,\text{ln}(r-r_0)$ as a best-fitting curve.  This function causes 
$e^{\Lambda(r)}$ to go to infinity relatively slowly, producing a small 
proper distance, as exemplified by Eq.~(\ref{E:proper2}).  So for small 
$|\Phi'(r)|$ we are dealing with a situation that can hardly be 
improved.  As $|\Phi'(r)|$ increases, however, the ln-functions used so 
far produce an ever poorer fit.  The question is whether this deviation 
from the ln-form can be quantified.  A related question might be: given 
the abundance of solutions and the resulting possibility of naturally 
occurring wormholes, can one estimate the likelihood of finding one that 
is actually traversable by humanoid travelers?

Suppose we consider the expansion 
\begin{equation}\label{E:log}
   \text{ln}(r-r_0)=\text{ln}\,r-\frac{r_0}{r}-\frac{1}{2}\frac{r_0^2}{r^2}
    -\frac{1}{3}\frac{r_0^3}{r^3}-\cdots
\end{equation}
obtained by expanding $(d/dr)\text{ln}(r-r_0)=1/(r-r_0)$ in a geometric 
series and then integrating.  Returning to Eq.~(\ref{E:proper2}), 
\begin{equation}\label{E:proper3}
  \ell(r)=\int_{r_0}^{kr_0}e^{1.20}e^{-0.23\left[\text{ln}\,r
   -\sum_{1}^{\infty}\frac{1}{n}\left(\frac{r_0}{r}\right)^n\right]}dr,
\end{equation}
which is equal to the value of $\ell(r)$ in Eq.~(\ref{E:proper2}).  To 
measure the deviation from this form, let us call $\Lambda(r)$ 
``log-like" if its expansion is
\[
  \text{ln}\,r-\frac{r_0}{r}-\frac{1}{2^p}\frac{r_0^2}{r^2}-\frac{1}{3^p}
     \frac{r_0^3}{r^3}-\cdots.
\]
(The condition $p<1$ makes the function ``less favorable" than the 
ln-function.) The series itself converges for all $r>r_0$ by the ratio 
test.  The proper distance now becomes
\begin{equation}\label{E:proper4}
  \ell(r)=e^a\int_{r_0}^{kr_0}e^{b\left[\text{ln}\,r-\sum_{1}^{\infty}
   \frac{1}{n^p}\left(\frac{r_0}{r}\right)^n\right]}dr.
\end{equation}
To study the behavior of $\ell(r)$ as a function of $p$, we let $r_0=4$, 
$k=2$, $a=1$, and $b=-0.20$.  To show how $\ell(r)$ drifts away 
from the original values (corresponding to $p=1$), $1/\ell(r)$ is plotted 
against $p$, as shown in Fig.~1.  The graph resembles the right half of a
\begin{figure}[htbp]
\begin{center}
\includegraphics[clip=true, draft=false, bb=0 0 299 212, angle=0, width=3.1in, height=3.1in, 
   viewport=50 50 220 200]{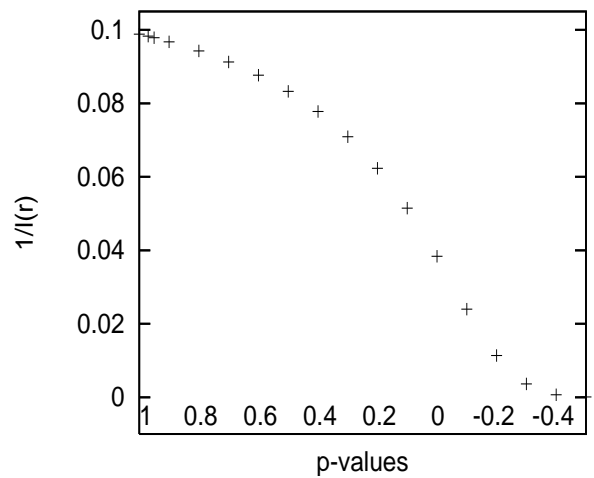}
\end{center}
\caption{\label{fig:figure2}A measure of traversability.}
\end{figure} 
normal curve.  So if we regard $1/\ell(r)$ as a measure of traversability, 
then, according to Fig.~1, the likelihood of encountering a naturally
occurring GCG wormhole that is also traversable may be higher than one should  
expect, perhaps even as high as 25\% (corresponding roughly to 
$1\ge p\ge 0.6$).

This outcome is fairly typical: increasing the throat size $r_0$ has 
little effect on the shape of the graph, while the values of $k$ and 
$a$ have no effect at all.  As $b$ decreases toward $-0.5$, the graph 
falls off more rapidly, but the percentage of traversable wormholes 
appears to remain between 10\% and 20\%.  Fortunately, even the 
worst-fitting cases have led to values between $-0.35$ and $0$.

So far we have assumed that $\alpha$ in the equation of state 
$p=-A/\rho^{\alpha}$ is close to unity, thereby producing 
the best-fitting curves.  As $\alpha$ decreases, one sees the same kind of 
falling off in the measure of traversability that appears in Fig. 1.  So 
the best chance for obtaining a traversable wormhole is a GCG that is close 
to the original Chaplygin gas.  However, this falling-off behavior can be 
partially compensated for by decreasing the constant $A$ in the equation 
of state. This can also be seen from Eq.~(\ref{E:basiceq3}): an 
increasing $1/\alpha$ is compensated for by decreasing $A$.  So a 
traversable wormhole could in principle exist for small $\alpha$, 
provided that the size of $A$ can be controlled (for example, by an 
advanced civilization).  In that case, however, the inequality 
(\ref{E:constraint3}), $\rho<A^{1/(\alpha+1)}$, becomes a constraint 
on the wormhole material.

\subsection{Junction to an external spacetime}\noindent
Observe that any curve of the form 
$\Lambda(r)=a+b\,\text{ln}(r-r_0)$ with $a>0$ and $b<0$ will 
eventually cross the horizontal axis.  So our spacetime is not 
asymptotically flat, the same problem that occurs in Refs.~\cite{pK06b}
and \cite{fL06}.  We would like to join $\Lambda(r)$ smoothly to a positive 
curve that goes to zero, as in Ref.~\cite{pK06a}.  

Assume that the extended curve has the form $g(r)=K/r^n$, starting at 
some $r=r_1$.  We require that $g(r_1)=\Lambda(r_1)$ and $g'(r_1)=
\Lambda'(r_1)$.  Eliminating $K$ and $n$, we obtain  
\[
   g(r)=\Lambda(r_1)\left(\frac{r}{r_1}\right)^
           {r_1\Lambda'(r_1)/ \Lambda(r_1)}.
\]
Since we are concerned mainly with the region around the throat, let us opt 
for an early cut-off, say at $r=2.2$.  (Besides, according to 
Ref.~\cite{fL06}, $r$ lies in a fairly narrow range, which is also apparent 
from the numerical/graphical output.)  As an example, for the case
$\Phi'(r)\equiv 0$, we have $\Lambda(r)=1.20-0.23\,\text{ln}(r-2)$.  Then 
for $r=2.2$, 
\[
    g(r)=1.57\left(\frac{r}{2.2}\right)^{-1.61}.
\]

Joining $\Lambda(r)$ to a positive curve going to zero makes the spacetime 
asymptotically flat.

\vspace{12 pt.}
\section{Comparison to Lobo's solution}\label{S:Lobo}\noindent
The case $\Phi'(r)\equiv 0$ allows a comparison to Lobo's exact solution 
\cite{fL06}.  From $\Lambda(r)=1.20-0.23\,\text{ln}(r-2)$, we calculate
\[
  b_1(r)=r\left[1-e^{-2.4}(r-2)^{0.46}\right].
\]
Lobo obtains
\[
  b_2(r)=r_0\left[\frac{64}{3}\frac{A\pi^2}{r_0^2}
             (r^6-r_0^6)+1\right]^{1/2}.
\]
On the above interval $[2, 2.2]$, using the same $A$ and $r_0$, 
$b_1(r)=b_2(r)$ after rounding off to two significant figures.  

\section{Summary}\noindent
This paper extends the general analysis in Ref.~\cite{pK06b} to wormholes 
supported by generalized Chaplygin gas (GCG).  The function $\Lambda(r)$ 
in line element (\ref{E:line1}) is obtained numerically from the Einstein field 
equations and the equation of state.  The conclusions are:

\begin{enumerate}

\item The wormhole spacetime meets the flare-out conditions at the throat.

\item Qualitatively, $\Lambda(r)$ remains the same for a wide range of redshift 
functions $\Phi(r)$.  Compared to the phantom energy supported wormholes 
in Ref.~\cite{pK06b}, event horizons are much less likely to occur.

\item Redshift functions with relatively small $|\Phi'(r)|$ yield regression 
equations closely fitting the form $\Lambda(r)=a+b\,\text{ln}(r-r_0)$.

\item Item 3 implies the existence of macroscopic wormholes with low tidal 
forces and relatively short proper distances near the throat, making the 
wormholes traversable by humanoid travelers.  (This criterion does not 
include proper traversal times.)

\item The junction to an external solution produces a spacetime that is 
asymptotically flat.

\item It is assumed above that $\alpha$ is close to unity in the equation of 
state $p=-A/\rho^{\alpha}$.  A smaller $\alpha$ produces less favorable 
results but can be compensated for by a smaller $A$.  However, a smaller 
$A$ would tighten the constraint $\rho<A^{1/(\alpha+1)}$ on the wormhole.

\item The abundance of solutions suggests that GCG wormholes may occur 
naturally.  A rough measure of traversability implies that the chances of 
finding one that is traversable by humanoid travelers may be fairly 
good.

\end{enumerate}

\end{document}